\documentclass{article}

\usepackage{PRIMEarxiv}

\usepackage[utf8]{inputenc} 
\usepackage[T1]{fontenc}    
\usepackage{hyperref}       
\usepackage{url}            
\usepackage{booktabs}       
\usepackage{amsfonts}       
\usepackage{nicefrac}       
\usepackage{microtype}      
\usepackage{lipsum}
\usepackage{fancyhdr}       
\usepackage{graphicx}       
\graphicspath{{media/}}     
\usepackage{amssymb}
\usepackage{cite}
\usepackage{eucal}
\usepackage{graphicx, multicol, latexsym, amsmath}
\usepackage{graphics}
\usepackage{blindtext}
\usepackage{subfigure}
\usepackage{caption}
\usepackage{capt-of}
\usepackage[noend]{algpseudocode}
\usepackage[boxruled]{algorithm2e}
\usepackage{enumitem}
\usepackage{multirow}
\usepackage{tabularx}
\usepackage{adjustbox}
\usepackage{float}

\title{FT-EALU: Fault Tolerant Arithmetic and Logic Unit for Critical Embedded and Real time Systems}

\author{
  Athena Abdi, Sina Shahoveisi \\
  Faculty of Computer Engineering, \\
  K. N. Toosi University of Technology, \\
  Tehran, Iran \\
  \texttt{a$\_$abdi@kntu.ac.ir, shahoveisysina@email.kntu.ac.ir} \\
}

\begin{document}
\maketitle

\begin{abstract}
In this paper, a fault-tolerant approach to mitigate transient and permanent faults of arithmetic and logic operations of embedded processors called FT-EALU is proposed. In this method, each operation is replicated in time and the derived final results are voted to generate the final output. To consider the effect of permanent faults, replicating identical operations in time is not sufficient, and diversifying the operands is required. To this aim in FT-EALU, we consider three distinct versions of input data and apply the target operation to them serially in time. To avoid high time overhead, we employ simple operators such as shift and swap to make an appropriate diversion in input data. Our proposed fault tolerance approach passes the replicated and diverse results to a novel weighted voter that is designed based on the reward/punishment strategy. For each version of execution, based on the proposed weighting approach a corresponding weight according to its correction capability confronting several faulty scenarios is defined. This weight defines the reliability of the result of each version of execution and determines its effect on the final result. The final result is generated bit by bit based on the weight of each execution and its computed result. These weights are determined statically through a design-time learning scheme according to applying several types of faults on various data bits. Based on the capability of execution versions on mitigating the permanent faults, positive or negative scores are assigned to them. These scores are integrated for several cases and normalized to derive the appropriate weight of each execution at bit level. These weights are employed as the predefined values in our proposed weighted voting approach as the trust factor of each bit of results to improve system reliability.
Several experiments are performed to show the efficiency of our proposed approach and based on them, FT-EALU is capable of correcting about $84.93\%$ and $69.71\%$ of permanent injected faults on single and double bits of input data.
\end{abstract}

\keywords{Fault tolerance system \and Permanent and transient faults \and Embedded systems \and Time redundancy \and Arithmetic and logic operations}

\section{Introduction}
\label{intro}
Embedded and real-time systems play an important role in modern applications. These systems are utilized in many critical applications such as health care, automation, avionic, communication, and so on. Embedded processors as the main part of these systems should perform correctly and autonomously~\cite{wang2017real,li2003real}. In many applications, the target system is employed in critical or hardly available situations where their correct and self-determined operation is very effective on their final performance. Arithmetic and logic operations are the main part of each processing unit and play a critical role in the system's functionality and reliability. Considering fault tolerance property in arithmetic and logic unit of embedded processors, make their operations more precise and autonomous~\cite{FT-ALU-2009,FAZELI2011-emalu,lizard}.

Along with the technology advances and shrink in size of transistors, the vulnerability of systems against transient and permanent faults increases. These faults affect various aspects of systems and make their results erroneous. Computation and arithmetic units are very sensitive to the faults because their results propagate to the system output, directly. Applying fault tolerance approaches in the computation unit is necessary to meet its correctness and autonomous requirements~\cite{xia2021error,ft-alu-2012,lizard}. Redundancy is the main solution in designing fault-tolerant systems and could be applied in various aspects of the system. Depending on the employed level of redundancy, fault detection or correction is possible in the cost of various performance and area overheads. In the case of transient faults with a very short lifetime, detection could be solely enough that leads to system restarting and recovery. Oppositely, permanent faults are more complicated and remain at the systems for a longer time. Thus, considering more complicated approaches that include correction and recovery along with detection methods are essential for their mitigation~\cite{ft-redundancy-2021,Tay2017,nelson1990fault,abdi2012}.

Time redundancy is an efficient fault-tolerance approach that replicates the system execution in time. In this approach, multiple replicated versions of the system are considered and repeat the same operation, serially. By replicating the execution of operands and comparing the results, transient faults are simply tolerated. To consider permanent faults in these methods, making variations in each replicated version of execution is beneficial and facilitates fault detection by comparing the differentiated results of computation. Detection capability is not enough for making a system fault-tolerant and locating the faulty place and correcting it consequently, are also required. These capabilities are achieved through considering extra controlling schemes such as reconfiguration or voting.
Time redundancy-based methods are efficient in terms of hardware cost compared to parallel execution and are considered as a proper choice for designing fault-tolerant embedded and real-time systems due to their time predictability~\cite{nelson1990fault,wang2017real,abdi2018hystery,alvarez2018mixing}.

Since in embedded and real-time systems the execution time analysis is performed based on worst case estimation, applying some time overhead during system execution could be acceptable. Normal executions are generally shorter than the worst-case estimation, so the saved time could be employed for the replicated executions and making the system fault-tolerant. The main mission of embedded and real-time systems could be summarized in control applications where the exact and correct computations are very important. Computations are mainly performed in the arithmetic and logic unit (ALU) of the processors so designing them fault-tolerant is very important. In this context, applying redundancy to various aspects of the system is applicable but the resulting overhead should be studied precisely based on the requirements of embedded systems~\cite{FTadder-MEJ,berger-Tabriz,acharya2018berger-iop,barredo2020efficiency}.

In this paper, we propose our fault-tolerant time redundancy-based ALU architecture for embedded and real-time systems (FT-EALU). The main purpose of this method is to mitigate the transient and permanent faults that occurred in the arithmetic and logic unit of embedded processors. Since most embedded and real-time systems are autonomous, their self-recovery from potential faults is required. In this context, the passive fault tolerance approaches that perform voting among the replicated versions of the operation and do not require additional management are convenient. Transient faults are detected by comparing the results of replicated versions of computation and generally corrected as the result of performing a simple voting mechanism among them due to their short remaining time in the system.
Reversely, to correct permanent faults, repeating the execution is not enough due to their long lifetime, and mitigation approaches including fault locating and recovery are also required. Moreover, by replicating the same version of the execution, permanent faults will not be detected, and applying variations to each version is necessary to make a diversion in the final results.
In our proposed approach, simple and low overhead modifications of input data such as shift and swap are considered to avoid hardware costs and make sufficient diversity in input data to mitigate permanent faults. By diversifying the various versions of computation, different results, due to the effect of permanent faults, are generated that could be mitigated during the voting process. Traditional majority voting schemes are generally not efficient for mitigating permanent and inclusive faults so employing an appropriate weighted voter that identifies the trusted results and magnifies their effect on final output improves the efficiency.
Experimental results show that our proposed FT-EALU method is capable of correcting $84.93\%$ of injected permanent faults in arithmetic and logic operations.
The main contributions of our proposed fault-tolerant approach could be summarized as follows:

\begin{itemize}
  \item Presenting a time redundancy-based fault-tolerant approach to correct transient and permanent faults of computation in arithmetic and logic unit of embedded systems;
  \item Presenting an adaptive and software-implemented weighted voting approach based on reward and punishment trust coefficient factor to cover the effect of faults and isolate them from final result;
  \item Employing a learning-based approach to define the appropriate weight for each bit of the results in voting process based on various fault injection scenarios to limit the cost and make the proposed method more autonomous;
  \item Presenting an adaptive and self-tuned architecture to adjust the system detection and correction capability proportional to its tolerable overhead.
\end{itemize}

The rest of this paper is organized as follows: The related works are reviewed in section 2. In Section 3, the details of our proposed fault tolerance architecture including the detection and correction approaches are explained. The experimental results and evaluation of FT-ALU are provided in section 4. Finally, the conclusion remarks are mentioned in section 5.

\section{Related Work}
\label{related}
The arithmetic and logic unit is one of the most critical elements of embedded systems that directly affect the correctness and reliability of the system output~\cite{lizard,FTadder-MEJ}. The fault-tolerant design of this component is very important and effective on system reliability. Based on the applications of embedded systems, their reliability and autonomous operation are very important. In this context, several studies concentrate on fault-tolerant computation and ALU operation but at different levels and aspects. The realization of fault tolerance in system design is generally reflected by redundancy in its different aspects. In the design of fault-tolerant ALU, mainly hardware and information-based redundancy approaches are proposed~\cite{lizard,berger-Tabriz,akbar2020self-repair-adder}.

Information redundancy-based methods add error detection/correction codes to data to protect them against potential faults. Here, encoding and decoding modules are required because, in ALU, the focus is on computation so the value of input data is important and should be unchanged to have the correct result. Various methods based on parity, residue, and Berger codes are proposed in this category.
In~\cite{acharya2018berger-iop} a fault-tolerant architecture for embedded processors based on employing Berger code is proposed. This method mainly tolerates transient faults and expands the operands by a logarithmic factor of their length. Extracting the raw and intended output from its encoded version require decoding hardware that leads to hardware overhead which is hardly acceptable in embedded and real-time systems due to their limitations. In~\cite{parity-2003-TVLSI}, a fault-tolerant architecture for ALU and adder based on adding parity codes is presented. In this approach, the hardware cost caused by parity codes is low but the fault handling approach requires considering an extra ALU which is not applicable in most embedded systems. Generally, information redundancy-based methods are not appropriate for arithmetic circuits because it is probable that the added codes modify during the computation in ALU and lose their efficiency.

In~\cite{towhidy2019-residu-jcsc}, a fault-tolerant approach based on residue codes is proposed. In this study to overcome the high hardware overhead of previous research, a new residue checker circuit is proposed by considering more compression. However, the final hardware overhead decreases but still it is high for embedded systems. In~\cite{thakral2020novel-parity-id}, to make the ALU fault-tolerant, parity gates are employed which have limited capability in fault tolerance and are suitable to detect single-bit faults. Hardware redundancy-based methods enforce extra hardware to the embedded systems that are not acceptable in many applications where the area and performance of these systems are critical.
In~\cite{berger-Tabriz}, a fault-tolerant design of ALU using enhance Berger code is presented. The focus of this approach is on reducing the hardware and performance overhead of Berger code and its effectiveness in mitigating the permanent faults of ALU is studied.

In~\cite{FAZELI2011-emalu}, a fault-tolerant ALU architecture for embedded systems is proposed. In this approach, the unused sections of the ALU system are utilized for redundant computations and make the system fault-tolerant. This method shows good performance in terms of area and fault correction abilities but still adds extra hardware and delay to the system.
In~\cite{lizard}, a narrow width value-based method for handling hard faults of ALU is presented. In this method, two half-word ALUs are employed to detect system faults. After fault detection, each ALU is partitioned to locate and isolate the occurred fault. This approach uses the unused parts of the system to perform operations in parallel but its detection and isolation capabilities are limited based on the system capacity.

Employing duplication with comparison (DWC) and triple duplex redundancy (TMR) are other effective approaches for fault detection and correction and these methods are employed along with various re-execution approaches. Employing these methods in hardware redundancy-based approaches leads to high overhead, power consumption, and delay which is not acceptable in embedded and real-time systems. Time redundancy techniques decrease the hardware cost but prolong the execution time that could be managed in most real-time applications due to their WCET estimation. Convenient time redundancy methods such as RESO, RESWO, and REDWC~\cite{RESO,RESWO,REDWC} consider various modifications in different versions of computation to make them different and improve the detection capability.
In~\cite{gade2020run}, a TMR-based fault tolerance method based on considering the basic time redundancy approaches such as RESO \cite{RESO} is presented. The considered voting approach of this study is designed based on majority voter.
These methods are effective in fault detection at low cost and overhead but their coverage is improvable. Considering the combinations and enhanced versions of these techniques could be effective in the design of fault-tolerant computing systems.

\section{Proposed Method}
\label{proposed method}
Design of reliable and autonomous embedded and real-time systems is very important due to their wide applications. Fault tolerance approaches are defined at different levels of system design and manage the occurred errors and keep them away from system output. Fault tolerance realizes along with active or passive redundancy in the system. Active fault tolerance approaches have fault detection, localization, and isolation/correction phases. Implementation of active methods make some delays in the system due to the required system reconfiguration. Oppositely, passive methods cover the fault from system output and do not make any interrupt in its operation. Our proposed fault-tolerant architecture is based on passive approaches and voting among multiple versions of operations. As it has been explained in previous sections, time redundancy-based approaches are the proper choice for embedded and real-time systems due to their worst-case time analysis during design that generally leads to some slacks in runtime execution. Our proposed method hardens the ALU against transient and permanent faults by employing enhanced time redundancy-based approaches and performing fair voting among them to isolate the occurred errors from system output and is proper for embedded systems.

Fault detection is the main prerequisite of the fault tolerance process. The better this step is performed, the fault tolerance method will be more efficient. Conventional time redundancy-based fault detection approaches, diversify the operations by simple modifications to enhance their detection capability confronting the permanent faults. These simple modifications include complementing the operands, shifting or swapping them that cause to make difference between original and replicated executions' results~\cite{RESO,RESWO,REDWC}. The fault detection coverage of these methods is further enhanced by modifying their traditional approaches~\cite{ICEEabdi,gade2020run,kermani2016}. In our proposed fault-tolerant approach, we utilize the original and modified versions of input data based on the mentioned enhanced methods as the system's basic blocks.

As it has been explained, considering effective detection approaches and making more simple differences among replicated executions, facilitates the fault correction and recovery process. Our proposed method employs multiple versions of execution with diversifying operands and vote among them. Thus, in our proposed approach basically, three various versions of input data are considered for required serial operations: a) raw operands b) left-shifted operands~\cite{RESO} and c) left-shifted and swapped operands~\cite{ICEEabdi}. Then, the generated results of three versions of executions are adapted. In this way, the result of the second execution is shifted back to the right and the result of the last operation is shifted to the right and swapped again to have the final results correctly.
To improve the detection capability, in our proposed approach the potential carry of arithmetic operations is also considered and adjusted after modifications. Using simple operations such as shift and swap, enforces less overhead to the system but distributes the effect of stuck-at faults and makes their detection very probable. It is possible to consider different variations on input data but based on our experiments, considering the selected three versions of execution leads to the most proper correction capability. Since their final results have less similarity to each other and in this way more shuffling is applied to input data that makes the effect of occurred permanent fault more clear in voting. The effect of other variations of input data on final output is studied in detail and reported in section~\ref{results}.

\begin{figure}
  \centering
  \includegraphics[width=4.5 in]{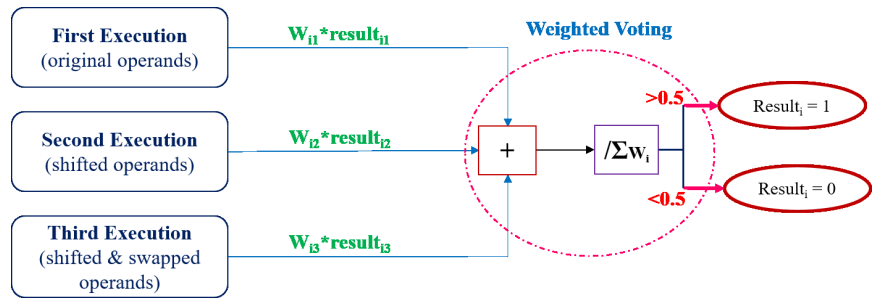}
  \caption{The high-level architecture of our proposed time redundancy-based fault tolerant approach}
  \label{fig-arch}
\end{figure}

To cover the effect of permanent faults that are resistant to repeated executions from the final output, voting among the results of defined operations on various input data is required. Majority voting is the simplest choice but its efficiency is limited due to the probability of wrong result repetition and its leading to final output. To overcome this, weighted voting is an appropriate solution but its efficiency is tightly dependent on the determined weights.
To specify the weight of each version of execution in our proposed method, we employ a \textit{trust factor} parameter based on the correction coverage of each version of execution. In this context, we propose two efficient weighting approaches based on reward/punishment, depending on the function of execution operation confronting the injected permanent faults. The first weighting approach is punitive and the second one employs the combination of reward and punishment in the evaluation and determining the \textit{trust factor}.
It should be noted that in our proposed approach, determining the weights is performed statically and through an offline learning approach. During this process, the correction capability of operands facing the injected permanent faults is investigated and based on their efficiency, a score is assigned to them. This score is derived based on several injected faults and determines the final weight of each operand in various execution versions. Since this process is performed statically and during system design, it enforces no complexity and time overhead. Afterward, the derived weights are utilized in the voting process among the results of execution versions.
In the proposed approach, the granularity of assigned weights is considered as fine as single bits to make it more precise.
The architecture of our proposed fault-tolerant approach is shown in figure~\ref{fig-arch}. In this figure, \textit{$W_{ij}$} shows the computed weight of $j^{th}$ version of execution for $i^{th}$ bit of the result (\textit{$result_{ij}$}). The various versions of execution in this figure, perform identical operations but on various operands that apply sequentially in time on a unique hardware platform. Employing various operands improves the permanent fault detection and correction capabilities, respectively.
The details of determining the weights of each execution and the employed voting mechanism are explained in the following subsection.

\subsection{The Proposed Voter Architecture}
\label{proposed-voter}

To determine the weights of various executions in the voting process, we propose and study two strategies based on reward and punishment.
Our first proposed weighted approach determines the weight of each execution (\textit{trust factor}) based on a punitive strategy. Here, the detection capability of considered executions confronting stuck-at faults in different bits of operands is studied. Since the operations are performed on various forms of operands, the effect of the fault and its propagation on different bits of the data are not the same. Thus, we define the weight for each bit of operands in execution versions based on their ability to correct the injected stuck at faults.
In the punitive weighting approach, the score of ideal executions that cover the injected stuck at fault does not change and the one(s) that propagate the error get a negative score.
This score is to each bit of operand and for each fault scenario, the unit negative score ("-1") divides among all versions of operation that are incapable of correcting the injected fault.
This score is a representation of punishment that reduces the \textit{trust factor} and the effect of the unreliable executions in the final output, consequently. In the other words, the score of unreliable execution is more negative and has less impact on deriving the final result of the system.

The total weight of each bit of operands in every version of execution is computed from its assigned scores in various fault injection scenarios. In this context, the total score of every single bit of input data, from all fault injection scenarios, divides by the considered faulty cases and then normalized. The normalization process scales the weights in a specified range and it is possible to consider various normalization approaches for deriving the final weights.
Here, we study three popular normalization approaches that are compared in section~\ref{results} and among them standard normalization scheme had the best results. Since standard normalization provides more scattered weights, it separates the results properly and reflects their unique effect on voting.
The explained weighting approach is performed at design time and through a learning-based scheme. In this context, several injected faults are considered during the training phase to determine the derived weight more efficiently.

Afterward, for each execution version, each bit of the result multiplies to its corresponding weight that is determined through the punitive approach and passes to the voter.
In the voter, the received results from three versions of execution are added together bit by bit and then divides by the total weights of the target bit in various versions. Then the derived result of the defined weighted average scheme should be interpreted. Due to the normalization process and final division, the final output of the voter is in the range of "zero" to "one". To map this bit-level result to binary values, we consider a threshold based on the performed operation. In this context, if the output of voter is greater than or equal to "0.5", it will be considered as one, otherwise, its value will be specified as zero in the binary scale.
Here, the value that is greater than "0.5" represents the case in which the operations with more than half of the total weights of the system indicate that the final result should be one and vice versa.
Thus, based on their domination in weight and the defined trust factor parameter, the most reliable versions of execution determine the final output of the system. It should be noted that other values of threshold are also studied but based on the mentioned explanations, the selected bound leads to the best results.
The defined punitive weighting approach is somehow strict and assigns the scores are greedily in one direction. In this approach, the weights are determined independent of the complexity of occurred fault and the weight of reliable execution is always zero. Thus, the effect of reliable versions is always constant in the final output and is not determinant.

Our second weighting approach is more efficient and defined as a combination of reward and punishment. Here, the assigned scores are negative for faulty cases (punishment) and positive for the corrected ones (reward). Moreover, in this approach, the scores are dynamically adapted based on the complexity and hardness of the faulty scenarios. To this aim, the maximum score of each faulty scenario, "+1", divides among the execution versions that are capable of mitigating it. For instance, if only one of the considered versions of execution is capable of correcting the injected fault, its corresponding score will be the highest, "+1", but in the case of two or three simultaneous fault corrections, this score divides by two or three respectively. Thus, the scores are set proportionally to the reliability of executions and the complexity of faulty scenarios.
The same procedure is also considered for punishment, and the complicated scenarios with hardly correctable errors have less punishment. The positive or negative score of each operation shows its assigned reward or punishment based on its fault correction capability.

\begin{figure}[h]
  \centering
  \includegraphics[width=5 in]{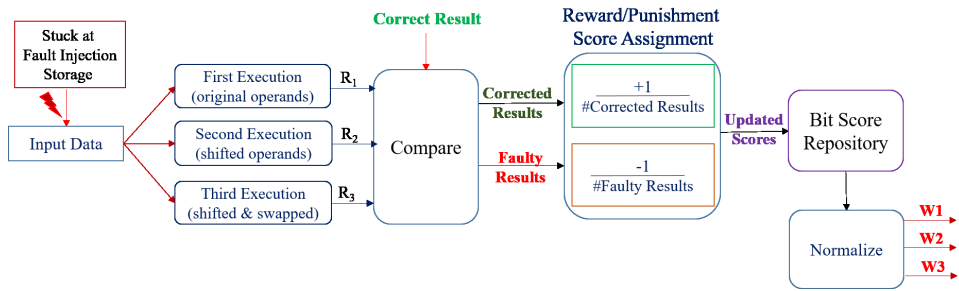}
  \caption{The details of our proposed weighting process based on a reward/punishment learning scheme through several fault injection scenarios}
  \label{fig_weight}
\end{figure}

Here again, after determining the scores of each bit, their corresponding weights are generated by normalization and division the same as the procedure that has been explained earlier for the first weighting approach.
The weights are determined for each bit of data to make the voting process more precise. These weights are extracted statically and at design time through a learning approach considering various permanent fault scenarios.
Figure \ref{fig_weight} shows the details of our proposed weighting approach. In this figure, the faulty scenarios are saved in fault injection storage and are applied to the input data during the training phase. Moreover, $"R_i"$ represents the intermediate results generated by the execution versions. At the end of the fault injection process, the final integrated scores are applied to the normalization box, and the weights are computed.

As this figure shows, based on comparing the results of each execution version to the correct output, its score is updated through a reward/punishment scheme. The reward of corrected results is determined as a positive score based on the complexity of the faulty situation. Similarly, the punishment is applied to the faulty results to decrease their score and affect the final output of the system. After applying all considered faults, the integrated final score of various executions is normalized and specified as the weight of each replicated operation.

It should be noted that the explained voting process is mainly implemented in software and enforces very limited complexity to hardware. Since the execution versions are distributed in time, they are implemented on the same hardware platform. Moreover, the weighted voting process is performed in software. However, if the hardware platform of the target system consists of redundant modules that could be utilized during the voting process, it will be appropriate and reduce the performance overhead. Since the target application of our proposed approach is embedded systems, extra hardware does not apply to them due to the cost and space constraints. Besides, extra hardware modules make the system more vulnerable to potential permanent faults. Thus, the hardware complexity of our proposed FT-EALU is very limited and could be expanded based on the requirements of the target system.
The details of the defined voting process and our proposed fault tolerance approach is presented in algorithm~\ref{alg_1}.

\begin{algorithm}[h]
\scriptsize
\caption{Pseudo-code of our proposed time redundancy-based fault tolerance approach (FT-EALU)}
\label{alg_1}

\textbf{// Replicate any operation three times in three various versions}

- Operation (F) on original input data (V1):

        $R\_V1$ = F(A,B)

- Operation (F) on left shifted input data (V2):

       AS = A \verb|<<| 1, BS = B \verb|<<| 1

       $RS\_V2$ = F(AS,BS)

       $R\_V2$ = $RS\_V2$ \verb|>>| 1

- Operation (F) on left shifted and swapped input data (V3):

       Halve the input data to AL, AH, BL, BH

       ALS = AL \verb|<<| 1, AHS = AH \verb|<<| 1, BLS = BL \verb|<<| 1, BHS = BH \verb|<<| 1

       $RHS\_V3$ = F(ALS,BLS) , $RLS\_V3$ = F(AHS,BHS)

       $RH\_V3$ = $RLS\_V3$ \verb|>>| 1 , $RL\_V3$ = $RHS\_V3$ \verb|>>| 1

       $R\_V3$ = ($RL\_V3 | RH\_V3$)

\textbf{// Perform the proposed weighted voting among the results}

\textit{- Determine the corresponding score of each bit based on the proposed reward-punishment approach}

       for t=1:N (number of fault injection scenarios)

       \quad for i=1:input data length

       \quad \quad if (three results are correct)

                 \quad \quad \quad $score(t)(V1_i)$ = $score(t)(V2_i)$ = $score(t)(V3_i)$ = $+\frac{1}{3}$;

        \quad \quad if (two results are correct (i.e V1,V2) and one is faulty (i,e V3))

                 \quad \quad \quad $score(t)(V1_i)$ = $score(t)(V2_i)$ = +0.5 , $score(t)(V3_i)$ = -1;

         \quad \quad if (two results are faulty (i.e V1,V2) and one is correct (i,e V3))

                 \quad \quad \quad $score(t)(V1_i)$ = $score(t)(V2_i)$ = -0.5 , $score(t)(V3_i)$ = +1;

          \quad \quad else

                 \quad \quad \quad $score(t)(V1_i)$ = $score(t)(V2_i)$ = $score(t)(V3_i)$ = $-\frac{1}{3}$;

        \quad end

       end

\textit{- Determine the normalized weight of each bit of input data in various versions of operation}

     $W_i(V1)$ = Standardize ( $\Sigma_t (score(t)(V1_i))/ N$ )

     $W_i(V2)$ = Standardize ($\Sigma_t (score(t)(V2_i))/ N$ )

     $W_i(V3)$ = Standardize ($\Sigma_t (score(t)(V3_i))/ N$ )

\textit{- Performing the weighted voting among various versions of operation to generate final output}

    for i=1:input data length

     \quad total $weight_i$ = ($W_i(V1)+W_i(V2)+W_i(V2)$)

     \quad Result (i) = [$W_i(V1)$ * $R\_V1_i$ + $W_i(V2)$ * $R\_V2_i$ + $W_i(V3)$ * $R\_V3_i$] / total $weight_i$

     \quad if (Result (i) $\geq$ 0.5)

     \quad \quad Final Result(i) = 1

     \quad else

     \quad \quad Final Result(i) = 0

    end

\end{algorithm}

\subsection{Illustrative Example}
\label{ill-example}

\begin{figure}[h]
  \centering
  \includegraphics[width=5.8 in]{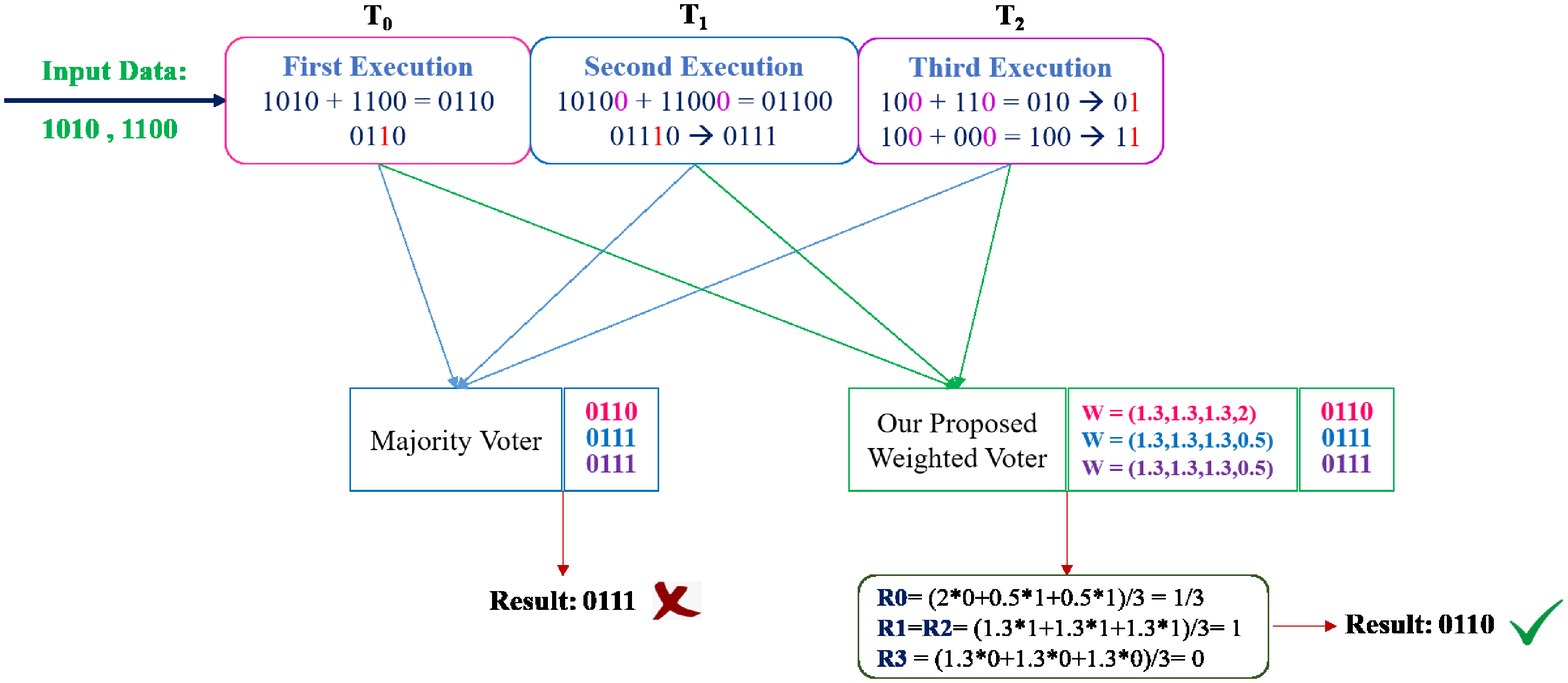}
  \caption{An illustrative example showing the details of our proposed fault tolerance approach on a tiny sample case}
  \label{fig-example}
\end{figure}

To show the workflow and details of our proposed fault-tolerance architecture (FT-EALU), in this section we apply it to an illustrative example. In this example, we focus on a four-bit input data with "stuck at one" permanent fault and the "add" operation as one of the key functions of ALU. As figure~\ref{fig-example} shows, in this example, the injected fault affects the second bit of data and makes it corrupted.
Here based on the explained approach, input data is applied to three distinct versions of operations: the original data, the shifted data, and the swapped-shifted data. Then the add operation is performed on the prepared input data and the replicated versions are executed in serial. The red bit in this figure represents the faulty bit and the result of each execution is passed to the voter, separately. The corresponding weight of each bit of data is prepared based on its capability in mitigating the injected fault.

In this example, the results of the majority and our proposed weighted voters are computed and compared. In the considered case, the final result of the second and third executions of this example are the same and the majority voter obeys them mistakenly.
Nevertheless in our proposed weighted voter, determining the score of each bit of results based on the defined trust factor leads to the correct result that covers the effect of injected fault, accurately.
In this example, the assigned weight of each bit of result is computed and normalized based on our proposed scoring algorithm and shown as \textit{W} vector in the figure. The final result is computed from performing a bitwise weighted sum on the assigned weights and results of each version of execution.
As this example shows, the overhead of our proposed fault tolerance approach is mainly related to the execution time of various versions of execution. Since the weighting process is mostly performed statically, it does not prolong the final execution time.
If the system hardware allows, modifying the input data for each version of execution could be performed in parallel with other executions. Eventually, this time overhead is managed by utilizing the extra time of the estimated deadline of real-time applications.

It is to be noted that, Our proposed \textit{FT-EALU} approach could be adaptively extended to five versions of operations and perform them voting among them like a 5MR system. In this case, the correction ability increases in the cost of more time overhead caused by employing more versions of execution. Adjusting the appropriate overhead and fault coverage levels are related to the application requirements and could be adaptively set by the designers based on the target system characteristics.

\section{Experimental Results}
\label{results}

In this section, to demonstrate the effectiveness of our proposed fault tolerance approach in making the ALU of embedded systems more reliable, several experiments and comparisons are presented.

\subsection{Simulation Setup}

To simulate and evaluate the proposed fault mitigation approach, our considered underlying platform is an ALU that implements basic operations such as "and", "or", "xor", "not", "add" and "subtract". The considered input data length is 4 and 16-bit and a comprehensive set of stuck-at faults are injected into them. Since our target underlying platform is an embedded system, the considered data length is appropriate and close to real cases but it could be easily extended to any other input length and fault pattern based on the target application requirements.

\begin{figure}[h]
  \centering
  \includegraphics[width=5.8 in]{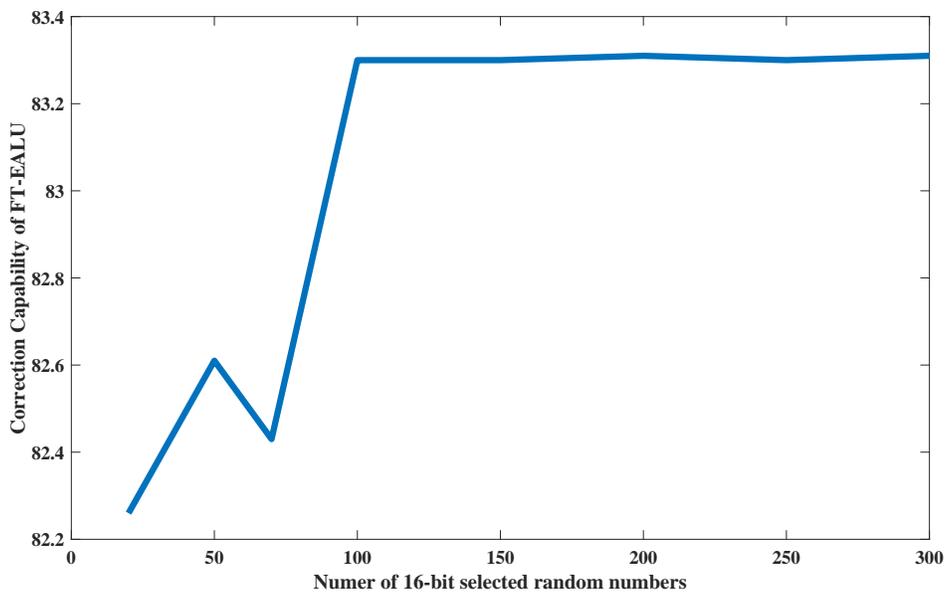}
  \caption{The correction capability of FT-EALU for different sizes of input data set selected from all permutations of 16-bit numbers}
  \label{fig-100-16bit}
\end{figure}

Our simulation platform implements the complete system architecture and focuses on permanent stuck-at faults that corrupt one or two bits of operands forever. The transient faults are fully mitigated during the time redundancy-based fault-tolerant approaches due to their short lifespan. Thus, we exclude transient faults from the experiments and focus on permanent ones. Since we study the effect of permanent faults on ALU and at the instructions level, they mainly affect input data or target operation. In both cases, the occurred fault modifies the value of data bits permanently. Stuck-at faults have a persistent effect on data bit and are proper to model the effect of permanent faults as they are employed in related research~\cite{FAZELI2011-emalu,ICEEabdi,gade2020run}.
In the following experiments, all permutations of single and double bit stuck at faults on 4-bit input data are considered. Moreover, all possible stuck-at faults (0 and 1) are applied to evaluate the proposed approach comprehensively.
Since considering all possible 16-bit numbers and applying all single and double bits stuck-at faults on them make the design space very large ($2^{21}$ various cases for single bit stuck at faults) which is not feasible in our simulation platform, we have limited them in our experiments. Here, selecting a limited number of random inputs is not representative so to make our experiments more precise and extended, we have selected one hundred random 16-bit numbers ten times (based on various randomness criteria) and applied all the possible stuck-at faults on them.
Our investigation and study show that in this case, one hundred random numbers is an appropriate choice that reflects the intended properties of the system. Figure \ref{fig-100-16bit} shows that by increasing the number of selected 16-bit numbers, the correction capability pattern of our proposed FT-EALU is not changed, and it nearly converges. The elbow point of this figure is in "one-hundred" point, and we consider it in our experiments. It should be noted that, due to applying all possible stuck-at faults on the considered data set, its size grows dramatically. Thus, the processing power of our underlying platform does not let us extend this study to more than 300 samples. However, the efficiency of our choice could be concluded from the final constant pattern of this figure.

As this figure shows, selecting one hundred 16-bit samples is representative and reflects the correction capability of our proposed method appropriately.
The proposed approach and the simulation platform are implemented by Python. In this context, a high-level model of ALU that is capable of performing the main logical and arithmetic operations is simulated and various combinational of faulty input data are applied to it. The simulated platform is implemented in Jupyter Notebook environment on an Intel Core i7 CPU with 8 GB RAM. During the simulation, we have implemented the considered operations of ALU, and some built-in modules such as math, random, and copy are utilized.

\subsection{Experiments and discussion}

To demonstrate the effectiveness of our proposed fault mitigation approach and compare it to the related research, in this section three distinct categories of experiments are considered as: (I) selecting the most proper variations of input data to enhance the correction capability of our proposed fault-tolerant approach by providing more variations in executive versions and cover the fault more efficiently, (II) evaluating the efficiency of our proposed voting approach in covering the permanent faults and determining the appropriate weights statically or permanently based on a learning-based strategy, (III) assessing the overhead of our proposed approach and comparing its correction capability to the related fault-tolerant approaches.

\textit{A. Selecting the proper combination of execution versions for the replicas}

The first experiment aims at determining the most proper combinations of input data variations that lead to the best correction coverage.
As it has been explained, in time redundancy-based approaches, it is required to make some variations in the replicated blocks of the system to make appropriate diversity to detect permanent faults.
Low-cost and lightweight operators are the best choice for making the required variations and improving the correction ability. However, the precise selection and combination of replicated versions are very important. Considering the similar operands in various versions of execution leads to repeated operations and analogous results that reduce the correction ability during the voting process.
To study the effect of combining different versions of execution in our proposed time redundancy-based approach, we evaluate it with various combinations of diversified operations. The modifications are mainly based on applying simple and low-cost operators such as shift, swap, and rotate on the input data of the operations.

In this context we consider seven distinct versions of operands for each operation: original input data, shifted input data (RESO)~\cite{RESO}, swapped input data (RESWO)~\cite{RESWO}, enhanced shifted input data (E-RESO)~\cite{ICEEabdi}, shifted and swapped input data (E-RESWO)~\cite{ICEEabdi} and two rotated input data approaches.
These modifications of operands in replicated executions are required to preserve and propagate the effect of injected faults in their corresponding results. In this way, the results of replicated executions are not similar and the occurred fault will be masked during the voting process.

In our TMR-based approach, the first version of execution is performed on original input data and the mentioned variations are applied to the replicated operations. Based on applying the mentioned candidate variations on input data of the replicated execution versions, six meaningful cases are formed. Here, applying similar variations such as shift and rotate in combinations is not considered due to their analogous results that reduce the correction ability.
The correction capability of the defined combinations of executions is presented and compared in table~\ref{table-exp1}.
The reported correction coverage of this table is computed in average for all considered arithmetic and logical operations of our simulated  ALU.

\begin{table}[h]
\caption {Investigating the correction capability of different combinations of variation in input data}
\centering
\begin{tabular}{|l|c|c|}
\hline
\multirow{2}{*}{\textbf{Correction Approach}} & \multicolumn{2}{|c|}{\textbf{Correction Coverage}} \\
\cline{2-3}
 & \textbf{4-bit} & \textbf{16-bit} \\ \hline
Basic operands $\vert$ RESO $\vert$ RESWO & 79.06\% & 80.17\% \\ \hline
Basic operands $\vert$ E-RESO $\vert$ E-RESWO & 60.28\% & 54.41\% \\ \hline
Basic operands $\vert$ Rotated RESO $\vert$ E-RESWO & 78.68\% & 80.93\% \\ \hline
Basic operands $\vert$ Rotated E-RESO $\vert$ E-RESWO & 43.99\% & 59.76\% \\ \hline
Basic operands $\vert$ Rotated E-RESO $\vert$ RESWO & 37.45\% & 32.33\% \\ \hline
Basic operands $\vert$ RESO $\vert$ E-RESWO & 80.35\% & 82.38\% \\ \hline
\end{tabular}
\label{table-exp1}
\end{table}

\begin{figure}
  \centering
  \includegraphics[width=5.8 in]{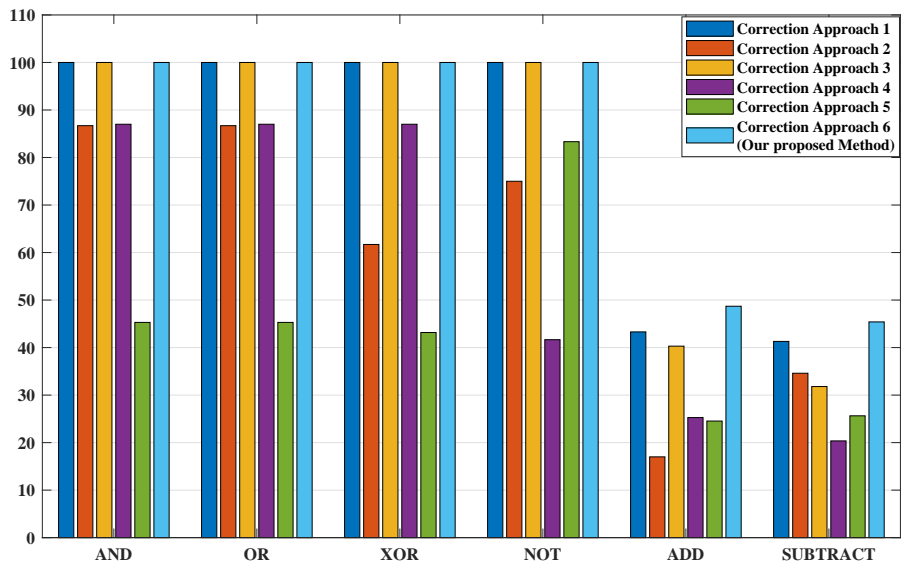}
  \caption{Correction capability of various combinations of operands categorized by ALU operations}
  \label{fig-exp1}
\end{figure}

As this table shows, combining different versions of executions in replicated versions lead to various correction capability.
Based on this table, combining the basic operands with their shifted (RESO) and shifted-swapped (E-RESWO) versions lead to the best correction coverage. Thus, our proposed FT-EALU considers this combination of execution that provides appropriate diversions in the results. As the result of this diversion, the effect of injected fault is kept in the intermediate results and then covered successfully during the voting process.
Moreover, based on the presented results, the correction coverage for 4 and 16-bit input data are almost matching which confirms the extension capability of this study.
It should be noted that the correction coverage is not the same for all ALU operations. The coverage of arithmetic operations is lower than logical ones due to their complexity.
Figure~\ref{fig-exp1} shows the correction capability of the studied combinations of table~\ref{table-exp1} categorized to different ALU operations. In this figure, the correction approaches are presented in the order of table~\ref{table-exp1}.

As this figure shows, the effectiveness of correction approaches in mitigating faults of logical operations is better than the arithmetic functions such as add and subtraction. Since arithmetic operations are more complex and potential of error propagation due to the carry and borrow transfer among data bits.
Our proposed fault-tolerance approach, considered as "Correction Approach 6" in this figure, shows the best coverage capability in all ALU operations due to its proper variations of input data and considering the effect of the carry bit on data swap and shifting.
It should be noted that the presented results of this figure are based on 4-bit data and the same pattern is valid for other data sizes based on our experiments.
Moreover, in this experiment, the basic majority voter is employed to show the effectiveness of various combinations of input data for different ALU operations.

\textit{B. Evaluation of the proposed weighted voter}

\begin{figure}
  \centering
  \includegraphics[width=5 in]{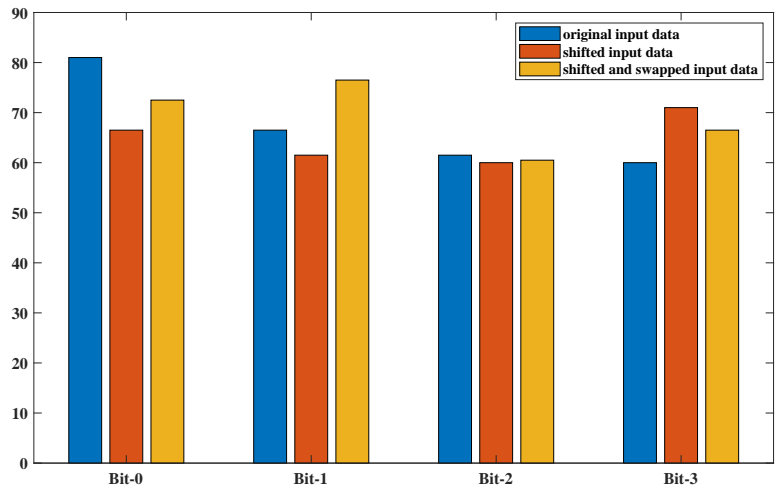}
  \caption{Robustness of different bits of various versions of input data against single and double stuck at faults during add operation (Bit 0: LSB and Bit 3: MSB)}
  \label{fig-exp2}
\end{figure}

The second experiment aims at demonstrating the effectiveness of our proposed weighted voting approach. Until now we have employed majority voting approach in our experiments.
Sometimes in complex cases, various versions of executions propagate the errors to the same position of system output, so considering majority voting is unreliable. Our proposed weighted voting approach evaluates the fault detection capability of various versions of computation and determines a weight for each of them statically. This weight represents the trust coefficient during the voting process and is determined per bit of input data. In this way the corresponding results of various versions of executions are not the same for the voter and their effect on final output is determined based on their trust factor.
In this experiment, the fault tolerance ability of conventional TMR that utilizes majority voters is compared to our proposed weighted voter's schemes. The weights of various executions in our proposed voter are determined statically through a learning-based approach. This weight represents their capability in error detection that is derived from their effectiveness in various cases.
As it has been explained in section~\ref{proposed-voter}, the first weighting approach is designed based on punishment of the incapable executions that propagate the errors and the second scheme employs reward and punishment schemes simultaneously. More capable approaches get higher weights and their corresponding result is more effective in final output due to their higher trustworthiness.
Figure~\ref{fig-exp2} shows the robustness of the considered execution versions in FT-EALU method separated by each bit of data for 4-bit input samples. In this figure, the average effect of single and double bit stuck at faults on input data during the add operation is studied. The effectiveness of various versions of execution in logical operations is very high and close to each other. Thus, to show the different trustworthiness of input bits more clearly, this experiment focuses on add operation which is more complicated.

Based on this figure, the robustness of different bits of input data against injected faults are not the same. Since some faults are covered during the operation and do not propagate to the output. These faults are not our target in our proposed fault-tolerant approach but they improve our system reliability. We try to focus on uncovered faults but the correction capability of each version of execution could be an appropriate guide for determining our target trust factor.
Based on figure~\ref{fig-exp2}, our considered execution versions in FT-EALU provide various robustness on input data and this could be utilized as the trust factor of each bit of result, laterally.
The reason behind this difference in robustness of input data bits is the effect of their employed variations. For instance, the most vulnerability of the shifted and swapped input data is in bit 2 that has the constant zero value due to shift operation. This constant value makes the data more vulnerable against stuck at faults.
The presented results of this experiment, verify that the reliability of different execution versions is not the same so applying them in the final result with various weights could be beneficial in fault correction.

As it has been explained, our weighted voting considers two different approaches to determine the emphasis of each version of execution in the final result. Moreover, during the weighting process, to have tuned final results, it is required to consider a proper normalization approach.
Linear normalization is the simplest approach for adjusting the computed weights of our proposed voters but we do not limit our study to it. We consider various efficient normalization approaches and study their effects on the correction capability of our proposed method.
In this experiment, the correction capability of our proposed voting approach in both proposed weighting schemes considering three different normalization methods: 1) linear (min-max) normalization, 2) adding the weight of each bit of data to the absolute minimum value of that bit in all experiments and 3) standard normalization are investigated.
The average results of this study for all operations of our considered ALU, are presented in table \ref{table-exp2}.

\begin{table}[h]
\caption {Correction capability of our proposed weighted voting approach considering the proposed punishment-based and punish-reward weighting approaches normalized by various normalization schemes}
\centering
\resizebox{\columnwidth}{!}{\begin{tabular}{l c c c c}
\hline
 & \multicolumn{3}{c}{\textbf{Correction Coverage}} \\ 
\cline{2-4}
 & \textbf{Normalization 1} & \textbf{Normalization 2}& \textbf{Normalization 3}  \\ \hline
Proposed voting scheme 1 (Punishment-based) & 81.76 \% & 78.22 \% & 82.57 \%   \\ \hline
Proposed voting scheme 2 (Reward-Punishment) & 82.37 \% &  80.23\% & 83.61\%	 \\ \hline
\end{tabular}}
\label{table-exp2}
\end{table}

As this table shows, employing the weighted voter and increasing the effect of more reliable executions in the final output, leads to better correction capability than majority voting.
As has been expected, the proposed punishment-based weighting approach is less efficient than the reward-punishment method. Since the punishment-based approach only reduces the effect of unreliable executions, unidirectional. However, the punish/reward weighting approach, adjust the weights in both incremental and decreasing sides and reinforces the effect of robust results in final outputs.

Until now, the corresponding weights of each execution are determined dynamically according to the system feedback. Since our target application is autonomous and real-time embedded systems, it is important to reduce performance overhead and estimate the weight of each operation statically. As it has been explained earlier, the weighting process could be more efficient and performed statically through a learning-based approach. To this aim, the corresponding and proper weights for each bit of various versions of execution are determined based on several random stuck-at faults.
To show the details of this learning approach and determine the appropriate weights for each bit of operands, in this experiment, a dataset including several randomly stuck at faults along with the errors that aim some vulnerable bits of data are considered.
Vulnerable bits are the ones that are at risk due to our considered variation techniques. For instance, the middle bits of data that are affected by the "swap" operations or the MSB and LSB that are modified by "shift" are considered vulnerable bits.
Propagating the stuck-at faults on these vulnerable bits to the final output is more probable and their correction is harder.
Our considered data set for the learning approach is formed of several random input data and various injected stuck at faults.
During the learning process, the data set is ten-folded and nine parts are employed for training and the remaining one fold is used to test the extracted values.
Table~\ref{table-exp3} shows the correction capability of FT-EALU based on the weights that are extracted through the described learned process. In this experiment, the weights are computed by the reward/punishment scheme due to its efficiency.
It should be noted that in this experiment, several random and specified stuck-at faults on single and double bits of 16-bit input data
are considered.
To perform the learning approach more efficiently, the dataset should be as large as possible and contain several faults. By considering more samples, the training phase of learning is implemented more efficiently. Thus, in this experiment, we expand our dataset to 400 random 16-bit data samples, and all permutations of single and double bits stuck-at faults are applied to it.
Moreover, in table~\ref{table-exp3}, the effect of defined normalization schemes are also studied.

\begin{table}[h]
\small
\caption{Correction capability of FT-EALU based on the static weights of voter determined by a ten-folding cross validation approach}
  \centering
  \begin{tabular}{c|c|c|c|c|c|c|}
    \cline{2-7}
     & \multicolumn{2}{c|}{\textbf{Normalization 1}} & \multicolumn{2}{c|}{\textbf{Normalization 2}} & \multicolumn{2}{c|}{\textbf{Normalization 3}}  \\
    \cline{2-7}
     & 1-bit SA & 2-bit SA & 1-bit SA & 2-bit SA & 1-bit SA & 2-bit SA \\ \hline
    \multicolumn{1}{|c|}{\textbf{FT-EALU}}  & 82.09\%  & 68.83\%  & 83.42\%  & 68.88\%  & 84.93\%  &  69.71\% \\
    \hline
  \end{tabular}
  \label{table-exp3}
\end{table}

As this table shows, the extracted weights provide appropriate correction coverage and are applicable to utilize as predefined weights in other unknown scenarios. Based on the results, the standard normalization has the best efficiency and could be integrated with our proposed weighting approach. These results are extracted based on various cases and could be simply utilized in other data lengths and various applications.

To demonstrate the efficiency of our proposed learning approach in determining appropriate weights, its correction capability is compared to the results of table~\ref{table-exp2} that is extracted the weights directly from the input data. By applying the extracted weights of learning to the input data of previous experiment, the correction capability differs about $0.51\%$. Thus our learning-based weighting approach could be efficiently employed in any application by its predefined weights. This process reduces the performance overhead of our proposed fault tolerance approach dramatically and makes it more applicable for real-time applications.

\textit{C. Comparison of FT-EALU to related approaches}

The last class of experiments compares the correction capability and performance overhead of our proposed fault tolerance approach to the related methods.
To this aim, the information redundancy-based method of~\cite{berger-Tabriz}, hardware reuse approach of~\cite{FAZELI2011-emalu} and TMR-based method of~\cite{gade2020run} that utilizes the basic time redundancy schemes are considered.
In~\cite{berger-Tabriz} the fault mitigation of ALU is performed through an enhanced version of Berger code that modifies the operations and operands to make the show of fault and tolerate it during the computation. In~\cite{FAZELI2011-emalu} a hardware redundancy-based approach to duplicate the operations is considered and its cost is moderated through the idle module reuse approach. The time redundancy-based approach of~\cite{gade2020run}, protects ALU through replicated operations in time and performs a majority voting among the results.
The comparison results of our proposed FT-EALU to these approaches in terms of correction capability and performance overhead are presented in table~\ref{table-exp4}.
Since the reported results of the considered related studies are mostly based on addition, in this experiment we have limited the ALU operation to "add". Moreover, the overhead of related studies is different due to their various redundancy schemes and. Hence, to have a fair and high-level comparison, we have not reimplemented the related methods and presented their reported correction capability and overhead.

\begin{table}
  \caption{Comparison of our proposed FT-EALU to related studies in terms of fault correction capability and performance overhead for "add" operation}
  \centering
  \resizebox{\columnwidth}{!}{\begin{tabular}{l|c|c}
    \hline
    Fault Tolerance Approach & Average Correction Coverage & Performance overhead over baseline\\
    \hline
    Information redundancy approach~\cite{berger-Tabriz} & 46.2\%  & 16.3\%  \\
    Hardware reuse approach~\cite{FAZELI2011-emalu} & 38.4\%  & 9.38\% \\
    TMR-based time redundancy approach~\cite{gade2020run} & 48.7\% & 35.99\% \\
    FT-EALU &  52.5\% & 37.74\% \\
    \hline
  \end{tabular}}
  \label{table-exp4}
\end{table}

As this table shows, the correction capability of our proposed FT-EALU outperforms the related approaches by $9.4\%$ on average. Since our research aims at critical embedded systems, their correction capability and robustness in presence of permanent faults are the main objectives. Besides the system constraints such as application deadline, cost and power consumption should be considered.
As it has been explained in section~\ref{proposed method}, to meet these constraints our proposed weighted voting unit is implemented in software and the weights are determined statically through a learning-based scheme. Moreover, the performance overhead of our proposed approach consists of triplicating each operation and performing a bit-wise voting among the results considering the determined weights of the learning approach.
Based on our implementation, the performance overhead of FT-EALU is about $37.74\%$ in cost of $84.93\%$ correction capability for all ALU operations which is reasonable in critical embedded systems that require high reliability.
This performance overhead is related to the replicated execution of various versions of operations in time and implementing the proposed weighted voter in software. If there exist redundant modules in the hardware platform of the target system, this cost could be reduced by implementing some of the mentioned operations in parallel by hardware.
Moreover, our proposed approach does not enforce extra hardware costs due to its replication in time. Since applying redundant hardware is not applicable in embedded systems and makes the system more vulnerable to faults, our proposed weighted voter is implemented in software.

Finally, the correction capability of our proposed approach could be improved by upgrading it to 5MR-based architecture. If the requirements of the target system in terms of timing constraints allow, it is possible to improve the correction capability of FT-EALU. In this context and based on our experiments, the correction capability of FT-EALU could be improved to  $87.3\%$ in cost of $68.59\%$ performance overhead. Thus, the existing trade-off between correction capability and performance overhead of FT-EALU could be adjusted based on the system requirements and the target application.

\section{Conclusion Remarks}

In this paper, a new fault-tolerance approach for ALU of embedded and real-time systems is presented. Due to the expanded applications of embedded systems, their precise and autonomous operation is critical. The arithmetic and logic unit (ALU) as the main part of the processor leads the basic operations of the system and affects its final output directly. Thus, our proposed fault tolerance approach (FT-EALU) aims at making the ALU reliable through a time redundancy-based scheme. FT-EALU mitigates transient faults by replicating the execution versions in time. Moreover, it covers permanent faults by applying variations in input data of replicated computations in time. The mentioned variations are applied on input data by simple modifications such as shift and swap that improve the correction capability along with forcing low-performance overhead to the system.
To mitigate the faults and determine the final result, FT-EALU performs weighted voting among the serial replicated executions. In this way, the effect of the occurred faults will be handled during the operation and autonomously as it is expected in embedded applications.

The efficiency of the considered weighted voting approach is tightly related to determining the weights. In FT-EALU, this weight is specified based on the fault mitigation capability of various versions of execution and through an adaptive reward/punishment approach. The weights are determined statically and during system design based on several fault injection scenarios. The extracted weights are extracted for each bit of data to improve the precision of our proposed approach. Then, these weights are applied to the results of the execution versions separately and the final output is generated by their integration.

Several experiments are performed to demonstrate the effectiveness of our proposed fault tolerance approach considering the main ALU operations and many fault injection cases. Experimental results show that our proposed approach is capable of correcting $84.93\%$ of the injected single bit stuck-at faults on arithmetic and logical operations on average. The performance overhead of FT-EALU is mainly related to the serial executions and could be decreased by reusing the redundant hardware of the system.
As a future trend, applying more low-cost variations on operations and decreasing the time overhead of the proposed method by effective approaches such as approximate computing is proposed. Moreover, considering a dynamic threshold to determine the final result of the proposed weighted voter by evolutionary schemes could be interesting.

\bibliographystyle{unsrt}  
\bibliography{references}

\end{document}